\documentclass[aps,prl,nofootinbib,preprintnumbers,floatfix,twocolumn,superscriptaddress]{revtex4}
\usepackage{pgfplots}
\usepackage{graphicx}
\usepackage{amssymb}
\usepackage{amsmath}
\usepackage{amsthm}
\usepackage{mathrsfs}
\usepackage{bm}
\usepackage{latexsym}
\usepackage{ulem}
\usepackage{textcomp}
\usepackage{color}
\usepackage{array, mathtools}
\usepackage{amsmath,amsfonts,amssymb}
\usepackage{tikz}
\usepackage{physics}
\usepackage{hyperref}
\hypersetup{
        colorlinks=true,  % Enables colored links
        linkcolor=blue,   % Color for internal links (e.g., table of contents, references)
        urlcolor=magenta,    % Color for external URLs (used with \url and \href)
        filecolor=black,% Color for links to local files
        citecolor=cyan,  % Color for citations
    }
\newcommand{\beq}{\begin{equation}} 
\newcommand{\eeq}{\end{equation}} 
\newcommand{\bea}{\begin{eqnarray}} 
\newcommand{\eea}{\end{eqnarray}} 
\def\benu{\begin{enumerate}}
\def\eenu{\end{enumerate}}
\def\nn{\nonumber}
\def\nn{\nonumber}

\def\pa{{\partial}}
\def\l{\left}
\def\r{\right}
\def\d{{\rm d}}
\def\f{\frac}

\def\R{\mathbb{R}}
\def\Z{\mathbb{Z}}
\def\t{\theta}
\def\p{\phi}
\def\a{\alpha}

\newcommand{\ie}{\textit{i.e.~}}
\newtheorem{definition}{Definition}
\begin{document}
%%%%%%%%%%%%%%%%%%%%%%%%%%%%%%%%%%%%%%%%%%%%%%%%%%%%%%%%%%%%%%%%%%%%%%%%%%%%%%%%%%%%%%%%%%%%%%%%%%%%%%%
\title{Magnetic flux and its topological effects in Aharonov-Bohm effect}
\author{Manvendra Somvanshi}
\affiliation{Department of Mathematical Sciences,
Indian Institute of Science Education and Research (IISER) Mohali, Knowledge City, Sector 81,
SAS Nagar, Mohali, Manauli PO, 140306, Punjab, India.}
\author{D. Jaffino Stargen}
\email{jaffinostargend@gmail.com}
\affiliation{Department of Physics, Global Academy of Technology,
Aditya Layout, RR Nagar, Bengaluru, Karnataka 560098, India}
%%%%%%%%%%%%%%%%%%%%%%%%%%%%%%%%%%%%%%%%%%%%%%%%%%%%%%%%%%%%%%%%%%%%%%%%%%%%%%%%%%%%%%%%%%%%%%%%%%%%%%%
\begin{abstract}
The Aharonov-Bohm effect is a physical phenomenon in which the quantum state of a charged particle
acquires a phase shift that is directly proportional to the magnetic flux, $\Phi$, due to a
(classical) magnetic field, ${\mathbf B}$, which is confined in a spatial region from which
the magnetic field cannot escape. Even though the charged particle
is not allowed to interact with the magnetic field, it accumulates a phase shift 
that affects the interference pattern produced. Not surprisingly, this apparent nonlocality
is puzzling and counter intuitive.
In this work, we provide an explanation that explains the physics underlying this apparent
nonlocality. We find that the role of the confined magnetic field is to impart
a puncture in the configuration space, $\mathbb{R}^2$, of the charge. Therefore, the
quantum state corresponding to the charged quantum particle acquires the phase shift due to its
response to the modified topology of the configuration space, $\mathbb{R}^2-\{0\}$,
corresponding to the charge.
\end{abstract}
\maketitle
%%%%%%%%%%%%%%%%%%%%%%%%%%%%%%%%%%%%%%%%%%%%%%%%%%%%%%%%%%%%%%%%%%%%%%%%%%%%%%%%%%%%%%%%%%%%%%%%%%%%%%%
\noindent
{\it {\bf Introduction}}:
In classical mechanics electromagnetic potentials are mere bookkeeping device that aids calculations,
and they do not manifest any physical effects; but in quantum mechanics they can produce
explicit physical effects. One such physical effect is known as Aharonov-Bohm effect,
where the quantum state of a charged quantum particle acquires a phase factor due to an external
electromagnetic field, interestingly, without the charged particle passing through the field \cite{AB1959}.
Specifically, as shown in Fig.(\ref{fig:AB}), in Aharonov-Bohm effect a beam of charged quantum
particles produce interference pattern
due to the presence of a constant magnetic flux that is passing through a region from which the magnetic
field is not allowed to escape.
%and the charged particles are rigorously excluded from the region.
Even though the charged particles do not interact with the magnetic field, they
respond to the magnetic field confined in the particular (shaded) region, and produce interference pattern
as per the magnetic flux produced by the magnetic field \cite{AB1959}.

It is well known that
the Hamiltonian corresponding to a charged particle of charge, $q$, in electromagnetic field is
\begin{align}\label{eqn:qHamiltonian}
 {\hat H} = \f{\l[{\hat {\bf p}}-q {\bf A}({\hat {\bf x}})\r]^2}{2m},
\end{align}
where ${\bf A}$ denotes the vector potential that describes the electromagnetic field \cite{Sakurai}.
Since the charged particles are rigorously
excluded from entering into the (shaded) region where magnetic field is nonzero and constant, one can assume
the vector potential, ${\bf A}$, in the region where the magnetic field is zero, to be
$A_{r}=0$, $A_{\theta}=\Phi/2\pi r$, and $A_{z}=0$, where $\Phi$ is a constant.
Therefore, the Hamiltonian corresponding to the charged particle in Eq.(\ref{eqn:qHamiltonian})
becomes \cite{AB1959}
\begin{align}\label{H-AB}
 {\hat H} = \f{-\hbar^2}{2m} \l\{\pdv{^2}{r^2}+\f{1}{r}\pdv{}{r}+\f{1}{r^2} \l(\pdv{}{\theta}
 + i\alpha\r)^2\r\},
\end{align}
where $$\alpha \equiv -\f{q\Phi}{2\pi}.$$
\begin{figure}\label{fig:AB}
    \centering
    \includegraphics[scale=0.21]{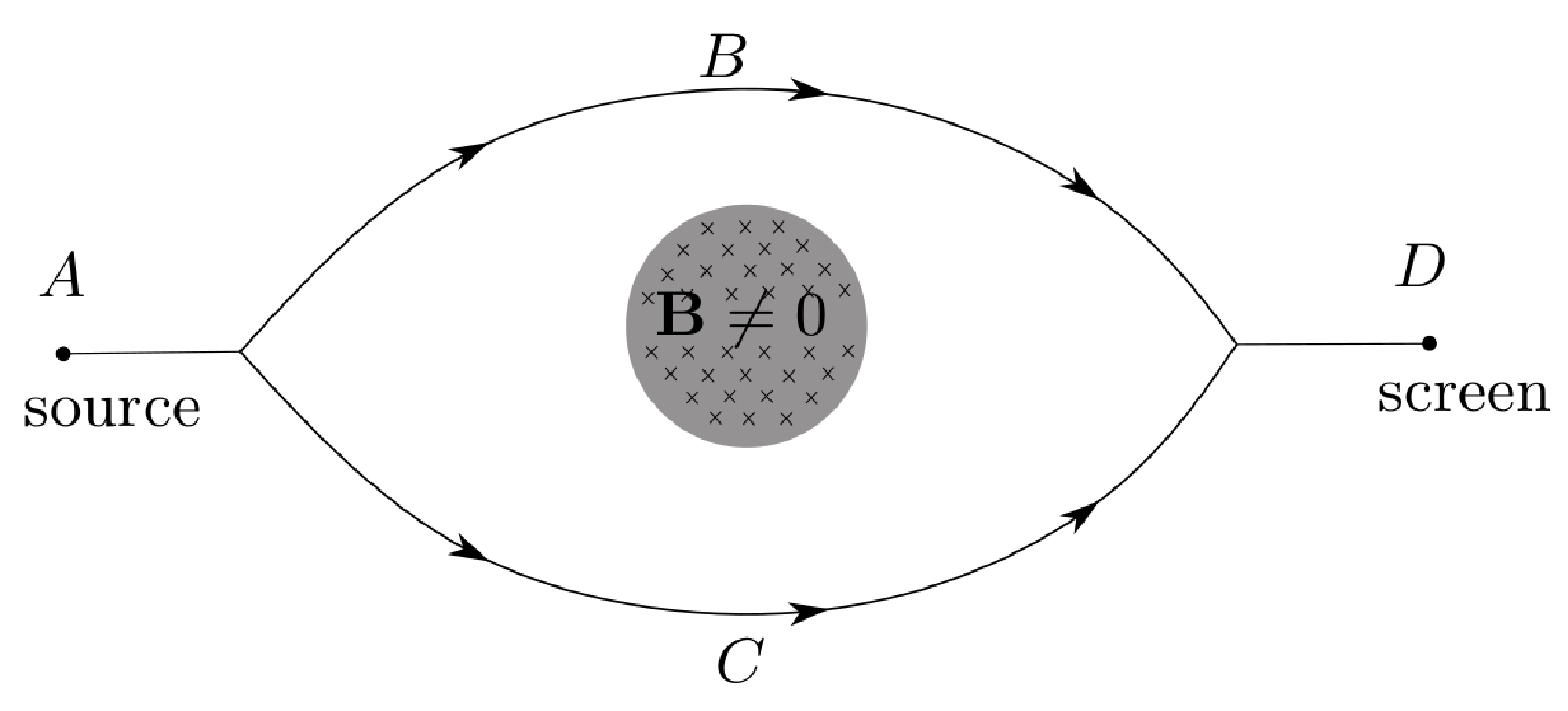}
    \caption{A schematic overview of the Aharonov-Bohm setup, where a charged quantum particle
    is allowed to move on a two dimensional plane, $\mathbb{R}^2$ (the configuration
    space of the particle); the shaded region represents a subset of the configuration space,
    $\mathbb{R}^2$, through which the magnetic field, $\mathbf{B}$, passes through; and
    the magnetic field is made vanishing everywhere outside the shaded region. In such a
    setup, a charged quantum particle can take two possible paths, ABD and ACD, and the
    phase difference between the quantum states of the charged particle corresponding to
    the paths, ABD and ACD, turns out to be proportional to charge, $q$, of the particle, and the
    magntic flux, $\Phi$, through the shaded region.}
    \label{fig:AB}
\end{figure}

If, as shown in Fig.(\ref{fig:AB}), $\psi_{\rm B}$ and $\psi_{\rm C}$ denote the wavefunctions of the
charged particles along the paths ABD and ACD, respectively, then the wavefunction, $\psi$,
at the point D is $\psi=\psi_{\rm B}+\psi_{\rm C}$. Therefore, the interference
pattern at the point D depends on the relative phase, $\Delta S$, between
the wavefunctions, $\psi_{\rm B}$ and $\psi_{\rm C}$, which is
\begin{align}
 \Delta S = q \oint {\bf A} \cdot \d{\bf x} = \f{q\Phi}{2\pi},
\end{align}
where the constant, $\Phi$, takes the form $\Phi=\int {\bf B} \cdot \d {\bf s}$, which is the magnetic
flux due to the non-zero magnetic field inside the shaded region \cite{AB1959}, as shown in Fig.(\ref{fig:AB}).
Therefore, even though the charged particle does not interact with the magnetic field, the effect of
the presence of magnetic field inside the isolated (shaded) region is explicit in the relative phase
shift, $\Delta S$, between the wavefunctions $\psi_{\rm B}$ and $\psi_{\rm C}$ \cite{AB1959}.
This ``nonlocal'' interaction between the magnetic field and the charged particle is indeed
interesting and puzzling. Many attempts were made in the literature to explain this
apparent nonlocality, but none has given a satisfactory explanation
\cite{Peshkin1981,Gonzalo1999,Kaufherr2004,Peshkin2010,Vaidman2012,Rohrlich2016,Vedral2020,
Cohen2023}.

Since the charged particles are excluded from entering into the region where magnetic field is
nonzero, one might tend to think that this exclusion modifies the topology of the configuration
space of the charged particle, and the existence of Aharonov-Bohm effect is due to this topological
change in the configuration space of the charged particle \cite{Zlotak2025}.
Even though the charged particle is excluded from a given region,
the phase factor acquired by the charged particle vanishes if the magnetic field in the above mentioned
region is zero. Therefore, the act of rigorously excluding the charged particle from a given region
does not manifest any physical effect, but the presence of a nonzero magnetic field in the concerned
region produces an explicit physical effect. This could motivate one to question whether the presence
of nonzero magnetic flux actually modifies the topology of the configuration space.

In this paper we describe Aharonov-Bohm effect as a phenomenon exhibited by a charged particle
due to the change in topology of the configuration space of the charged particle, which is imposed by
the magnetic flux passing through to the configuration space of the charged particle.
In particular, we investigate the quantum dynamics of a particle moving on a plane with a puncture
imposed on it. Due to the presence of puncture, the topology of the configuration space experienced
by the charged particle becomes nontrivial. Interestingly, the Hamiltonian operator obtained after
quantizing the plane with puncture becomes identical with the Hamiltonian operator of the charged particle moving on a plane, with
magnetic field lines piercing through the plane. This could imply that the presence
of magnetic flux passing through the configuration space, ${\mathbb R}^2$, of the charged particle modifies
its topology, and the emergence of Aharonov-Bohm effect is due to the {\it local} interaction of the
(charged) particle with the change in topology of its configuration space.
We quantize the punctured plane using Isham's group theoretic quantization procedure. Before proceeding
further to do quantum mechanics in the punctured plane, we briefly review the group theoretic
quantization below.
%%%%%%%%%%%%%%%%%%%%%%%%%%%%%%%%%%%%%%%%%%%%%%%%%%%%%%%%%%%%%%%%%%%%%%%%%%%%%%%%%%%%%%%%%%%%%%%%%%%%%%%
\vskip 5pt\noindent
{\bf Isham's group theoretic quantization}:
Canonical quantization is a popularly known quantization procedure in which classical variables
are associated to self-adjoint operators on a Hilbert space.
The vectors in the Hilbert space are called quantum states corresponding
to the quantized system. It is well known that canonical quantization works well in flat Euclidean space,
${\mathbb R}^n$, particularly due to the global vector space structure of ${\mathbb R}^n$, but it is
unsuitable for quantizing configuration spaces that possess nontrivial topology
\cite{von2018mathematical,wald1994quantum}. In such scenarios,
Isham's group theoretic quantization procedure becomes useful.

%Suppose, a classical dynamical system has a phase space that is a smooth symplectic manifold,
%$(M,\omega)$. The smooth functions, $f:M\to \R$, in the phase space, $(M,\omega)$, represent the
%classical observables of the dynamical system, and the time evolution of these classical
%observables is given by
%\begin{equation}
% \dot{f} = \{f,H\},
%\end{equation}
%where $H$ denotes the Hamiltonian, a smooth function on $M$, corresponding to the classical
%dynamical system, and $\{\cdot, \cdot\}$ signifies the Poisson bracket on the phase space, $(M,\omega)$.

Group theoretic quantization of a phase space, $(M,\omega)$, corresponding to a classical system
refers to a map $\hat{\cdot}:C\subset C^\infty(M) \to \mathscr{O}(\mathscr{H})$, where
$(M,\omega)$ is a smooth symplectic manifold, $\mathscr{O}(\mathscr{H})$ denotes the algebra of
self-adjoint linear operators on a Hilbert space $\mathscr{H}$, and $C$ is a subalgebra of
$C^\infty(M)$, such that the Poisson bracket algebra is related to the
commutator algebra in the following way
\begin{equation}\label{quant}
 \{f,g\} \mapsto -\f{i}{\hbar} [\hat{f}, \hat{g}].
\end{equation}
The group theoretic quantization procedure is primarily based on the
canonical group, $\mathscr{G}$, a Lie group corresponding to the phase space, $(M,\omega)$
\cite{isham}. The Lie algebra corresponding to
the canonical group, $\mathscr{G}$, bridges the Poisson bracket algebra of classical variables,
and the algebra of unitary operators on a Hilbert space, $\mathscr{H}$. The action
of elements of canonical group, $\mathscr{G}$, on the elements of the phase space, $(M,\omega)$,
induces a map between the Lie algebra of the canonical group, $\mathscr{G}$, and $C^\infty(M)$.
There are Lie group representations, $\mathcal{U}:G\to \text{Unitary}(\mathscr{H})$, that give a quantization
map from a subalgebra of $C^\infty(M)$ to operators in $\text{Unitary}(\mathscr{H})$.
%Note that the unitary
%representations of the canonical group, $\mathscr{G}$, may not be unique, and therefore the group
%theoretic quantization scheme assigns each unitary representation of the canonical group,
%$\mathscr{G}$, a particular quantization map.
An important remark about group theoretic
quantization is that the group product on the canonical group, $\mathscr{G}$, determines the
``Weyl-like" relations on the unitary operators on the Hilbert space, $\mathscr{H}$. Using these
weyl-like relations we determine the commutation relations between the operators on the Hilbert
space, $\mathscr{H}$, which is in contrast to the usual canonical quantization scheme employed for
quantizing ${\mathbb R}^{n}$ \cite{isham}.

To quantize a given phase space, $(M,\omega)$, employing group theoretic quantization procedure,
one has to perform the following steps below \cite{isham}:
\vskip 10pt\noindent
$\bullet$ \textbf{Step-I}:
\begin{definition}
 A smooth manifold, $M$, is said to be the homogeneous space of a Lie group, $G$, if the
 Lie group, $G$, admits a transitive action on $M$, i.e. $\forall x,y\in M$ there is a
 $g\in G$ such that $y=g\cdot x$.
\end{definition}
The first step is to construct a Lie group, $\mathscr{G}$, called the canonical group of the phase
space, $(M,\omega)$, such that the smooth manifold, $M$, is a Homogeneous space of $\mathscr{G}$, and the action of each
element of the group, $\mathscr{G}$, on the phase space manifold, $M$, is symplectomorphic.
\vskip 10pt\noindent
$\bullet$ \textbf{Step-II}: In this step, one constructs a map, $\gamma:\mathfrak{g} \to \text{VF}(M)$,
where $\mathfrak{g}$
denotes the Lie algebra corresponding to the canonical group, $\mathscr{G}$, and $\text{VF}(M)$
denotes the set of all vector fields in the smooth manifold, $M$.
%For this purpose, consider a map $\R\to \mathscr{G}$, that is given explicitly as
%$t \mapsto \exp(tA)$, where $A\in \mathfrak{g}$, and $\exp(tA) \in \mathscr{G}$. Therefore, the image of the map,
%$t \mapsto \exp(tA)$, is a one-parameter subgroup of $\mathscr{G}$, and
%
A vector field, $\gamma^A \in \text{VF}(M)$, is defined as
\begin{equation}
 \gamma^A_x(f) = \dv{f(\exp(-tA)x)}{t}\bigg|_{t=0}, 
\end{equation}
where $x\in M,\ f\in C^\infty( M)$, and $A \in \mathfrak{g}$.
%It can be shown straightforwardly that the map $\gamma:\mathfrak{g} \to \text{VF}(M)$ is a homomorphism
%(see Appendix.(\ref{app:C})), hence
%\begin{equation}
% [\gamma^A, \gamma^B] = \gamma^{[A,B]}.
%\end{equation}
The vector fields, $\gamma^A$, are called fundamental vector fields of the Lie algebra, $\mathfrak{g}$, corresponding
to the canonical group, $\mathscr{G}$.
\vskip 10pt\noindent
$\bullet$ \textbf{Step-III}:
\begin{definition}
 A vector field $\xi$ is said to be a Hamiltonian vector field, if there exists a function,
 $f\in C^\infty$(M), in the phase space, $(M,\omega)$, such that
 \begin{equation}
  \dd f(X) = \omega(\xi, X),\qq{where} X\in \text{VF(M)}.
 \end{equation}
\end{definition}
The goal in this step is to construct a homomorphism from the Lie algebra, $\mathfrak{g}$,
corresponding to the canonical group, $\mathscr{G}$, into the classical observables,
$f \in C^\infty( M)$. From Step-II above, we know that every element,
$A$, in the Lie algebra, $\mathfrak{g}$, corresponds to a fundamental vector field,
$\gamma^A \in \text{VF(M)}$. Moreover, every function, $f \in C^\infty( M)$, in the phase
space, $(M,\omega)$, corresponds to a Hamiltonian vector field, $\xi_{f}$.
Therefore, for the purpose of constructing a homomorphism from the Lie algebra, $\mathfrak{g}$,
into the classical observables, $C^\infty( M)$, we require the fundamental vector fields, $\gamma^{A}$, to be
Hamiltonian vector fields. Consequently, this restricts the canonical group, $\mathscr{G}$, to be such that all the
fundamental vector fields, $\gamma^A$, corresponding to the elements, $A$, in the Lie algebra,
$\mathfrak{g}$, to be
Hamiltonian vector fields.
%Supposing a canonical group, $\mathscr{G}$, satisfies this condition,
%then define a map $P:\mathfrak{g} \to C^\infty(M, \R)$, which is given by $A \mapsto P_A$, where the classical observable,
%$P_A$, corresponds to the field $-\gamma^{A}$.
%
\vskip 10pt\noindent
$\bullet$ \textbf{Step-IV}:
\begin{definition}
The action of a group, $G$, on a manifold, $M$, is said to be \textit{effective}, if any pair of elements $g_{1},g_{2}\in G$ are such that for all $x\in M$ we have $g_1x = g_2x$, then $g_1=g_2$.
\end{definition}
If two distinct elements of the canonical group, $\mathscr{G}$, acting on a point in the phase space,
$(M,\omega)$, does not lead to two distinct points in the phase space, then a Hamiltonian vector field
could correspond to more than one element in the Lie algebra, $\mathfrak{g}$.
This can be avoided if one restricts the action of the canonical group, $\mathscr{G}$, on the
phase space, $(M,\omega)$, to be effective.

\vskip 10pt\noindent
$\bullet$ \textbf{Step-V:}
%In general, the map, $P$, defined in Step-III may not be a homomorphism.
If $\xi_{f}$ be the Hamiltonian vector field corresponding to a classical observable,
$f \in C^{\infty}(M)$, then
\begin{equation}
 \xi_{P_{[A,B]}} = -\gamma^{[A, B]} = -[\gamma^A, \gamma^B] = -[\xi_A, \xi_B] = \xi_{\{P_A, P_B\}}.
\end{equation}
If $\xi_{f}=\xi_{g}$, then $f-g={\rm constant}$, then one can write
\begin{equation}
 P_{[A,B]} - \{P_A, P_B\} = z(A,B), \quad z(A,B)\in \R.
\end{equation}
Therefore, the map, $P$, is not a homomorphism in general. If $z(A,B)=0$, then the linear map, $P$, is an
algebra homomorphism into the Poisson algebra, $C^\infty(M)$, and is called a momentum map. In the cases
where the factor, $z(A,B)$, cannot be made to vanish by adding a constant to $P$, the Lie algebra is extended
to $\mathfrak{h} = \mathfrak{g} \oplus \R$, with the Lie bracket expressed as
\begin{equation}
 [(A,r), (B,s)] = ([A,B], z(A,B)).
\end{equation}
Consider the map, $P':\mathfrak{h}\to C^\infty(M)$, that is given by $P'_{(A,r)} = P_A + r$, where $r\in \R$,
then one can show that
\begin{eqnarray*}
 \{P'_{(A,r)}, P'_{(B,s)}\} &= \{P_A, P_B\} = P_{[A,B]} + z(A,B) \nn \\
 &= P'_{([A,B], z(A,B))} = P'_{[(A,r), (B,s)]},
\end{eqnarray*}
which means the map, $P'$, is a homomorphism. Moreover, the canonical group, $\mathscr{G}$, is simply replaced
by the unique simply connected Lie group, $H$, which has the Lie algebra, $\mathfrak{h}$.
\vskip 10pt\noindent
$\bullet$ \textbf{Step-VI:} Let $\mathscr{H}$ be a separable Hilbert space, on which let
$\text{Unitary}(\mathscr{H})$ be the group of unitary operators, and let the map
$U: \mathscr{G} \to \text{Unitary}(\mathscr{H})$ be a weakly continous, irreducible unitary representation of the
canonical group, $\mathscr{G}$, such that
\begin{equation}
 U(\exp(A)) = e^{-i\hat{K}_A},
\end{equation}
where $\hat{K}_A$ is a self adjoint operator on the Hilbert space, $\mathscr{H}$. Furthermore, since
\begin{align}
 & U((\exp(tA)\exp(sB))(\exp(-tA)\exp(-sB))) \nn \\
 &= \l(e^{-it\hat{K}_A}e^{-is\hat{K}_B}\r)\l(e^{it\hat{K}_A}e^{is\hat{K}_B}\r),
\end{align}
one obtains
\begin{equation}\label{eqn:KAKB}
 [\hat{K}_A, \hat{K}_B] = \hat{K}_{[A,B]}.
\end{equation}

As we know that the momentum map, $P:\mathfrak{g} \to C^{\infty}(M)$, is a Lie algebra homomorphism, and
Eq.(\ref{eqn:KAKB}) implies that a map $\hat{K}:\mathfrak{g} \to \text{SelfAdj}(\mathscr{H})$ is a homomorphism,
where $\text{SelfAdj}(\mathscr{H})$ is the algebra of Hermitian operators on the Hilbert space, $\mathscr{H}$,
the quantization map can be given by associating $P_A$ with $-i\hat{K}_A$. Therefore, it follows to
\begin{equation}
 \{P_A, P_B\} = P_{[A,B]} \mapsto -i\hat{K}_{[A,B]} = -i[\hat{K}_A, \hat{K}_B],
\end{equation}
which means the quantization map satisfies Eq.\eqref{quant}.
\vskip 5pt
\noindent
\vskip 5pt
\noindent
%%%%%%%%%%%%%%%%%%%%%%%%%%%%%%%%%%%%%%%%%%%%%%%%%%%%%%%%%%%%%%%%%%%%%%%%%%%%%%%%%%%%%%%%%%%%%%%%%%%%%%%
{\bf Quantum mechanics on the punctured plane}:
Employing Isham's group theoretic quantization scheme, as stated above, we quantize
a free particle of mass $m$ on punctured plane \cite{SuppMat}.
We find the momentum operators of the free particle in the position representation as
\begin{align}
 {\hat \pi}_{\phi} = -i\hbar \f{\pa}{\pa \phi} + \hbar \beta, ~~~~
 {\hat \pi}_{\rho} = -i\hbar \rho \f{\pa}{\pa \rho},
\end{align}
where $\beta$ is a real number which parametrizes the above operator representation,
and the angular operators in the position representation as \cite{SuppMat}
\begin{align}
 {\hat c} = \rho \cos\phi, ~~~~
 {\hat s} = \rho \sin\phi.
\end{align}
The operators, ${\hat \pi}_{\phi},~{\hat \pi}_{\rho},~{\hat c}$, and ${\hat s}$ satisfy the
commutation relations as below \cite{SuppMat}:
\begin{align}
 [{\hat s},{\hat \pi}_{\phi}]&=i{\hat c}, ~~~~
 [{\hat c},{\hat \pi}_{\phi}]=-i{\hat s}, \\
 [{\hat s},{\hat \pi}_{\rho}]&=i{\hat s}, ~~~~
 [{\hat c},{\hat \pi}_{\rho}]=i{\hat c}, \\
 [{\hat \pi}_{\phi},{\hat \pi}_{\rho}]&=0, ~~~~~
 [{\hat c},{\hat s}]=0.
\end{align}
%For more technical details regarding the quantization of free particle in the punctured plane,
%see Ref.\cite{MathPaper}.

Since the classical Hamiltonian, $H={\bf p}^2/2m$, of a free particle in the punctured plane in Cartesian
coordinates, $(x,p_x,y,p_y)$, can be written as
\begin{align}
 H = \f{(xp_x + yp_y)^2 + (xp_y - yp_x)^2}{2m(x^2+y^2)},
\end{align}
and since the Isham's group theoretic quantization leads to
\begin{align}
 x p_y-y p_x \mapsto \hat{\pi}_1, ~~ {\rm and} ~~ x p_x + y p_y \mapsto \hat{\pi}_2,
\end{align}
also 
\begin{align}
 x \mapsto \hat{c}, ~~ {\rm and} ~~ y \mapsto \hat{s},
\end{align}
the Hamiltonian operator, $\hat{H}$, for the free quantum particle in position representation
can be written as \cite{SuppMat}
\begin{align}\label{H-puncture}
 \hat{H} = -\f{\hbar^2}{2m} \l[\pdv[2]{}{\rho} + \f{1}{\rho} \pdv{}{\rho}
 + \f{1}{\rho^2}\l(\pdv{}{\phi} + i\beta \r)^2\r]. 
\end{align}
\vskip 5pt
\noindent
%%%%%%%%%%%%%%%%%%%%%%%%%%%%%%%%%%%%%%%%%%%%%%%%%%%%%%%%%%%%%%%%%%%%%%%%%%%%%%%%%%%%%%%%%%%%%%%%%%%%%%%
{\bf Does magnetic flux impart puncture in configuration space?}-- 
Note that the Hamiltonian in Eq.(\ref{H-AB}) corresponding to the charged particle in the
Aharonov-Bohm setup \cite{AB1959} in Fig.(\ref{fig:AB}) is identically
same as the Hamiltonian of a free particle in the punctured plane in Eq.(\ref{H-puncture}),
if the parameter, $\beta$, is assumed to be $\beta=\alpha \equiv -q\Phi/2\pi$.
The charge, $q$, and mass, $m$, of the particle in the
Aharonov-Bohm setup is not directly interacting with the magnetic field inside the shaded
region in the Aharonov-Bohm setup in Fig.(\ref{fig:AB}), since the magnetic field is confined and sealed
inside the shaded region \cite{AB1959}.
Therefore, it could be argued that
the charged particle responds only to the nontrivial
topology due to the puncture imparted by the magnetic flux, $\Phi$, that passes through the configuration
space, $\mathbb{R}^2$, of the charged particle. If the parameter, $\beta$, assumes this particular form:
$\beta=\alpha \equiv -q\Phi/2\pi$, then
the effect of charge, $q$, and magnetic flux, $\Phi$, on the quantum dynamics of the charged particle
is to choose a unitary representation in
the Hilbert space, ${\mathscr H}$, which is parametrized by the parameter, $\beta$. In other words,
the charged particle in the Aharonov-Bohm setup behaves like a free particle of mass, $m$,
moving in the plane with a puncture that is imparted by the magnetic flux, $\Phi$; and
the product of charge, $q$, of the particle, and the magnetic
flux, $\Phi$, chooses a particular unitary operator representation.
%%%%%%%%%%%%%%%%%%%%%%%%%%%%%%%%%%%%%%%%%%%%%%%%%%%%%%%%%%%%%%%%%%%%%%%%%%%%%%%%%%%%%%%%%%%%%%%%%%%%%%%
\vskip 5pt
\noindent
{\it {\bf Conclusions}}--
The Aharonov-Bohm effect is apparently a nonlocal phenomenon, where the quantum states of a charged particle
is affected by the magnetic field that is not directly interacting with the charged particle \cite{AB1959}.
Various
attempts were made to provide a suitable explanation underlying the apparent nonlocality, but none provides
a satisfactory explanation \cite{Peshkin1981,Gonzalo1999,Kaufherr2004,Peshkin2010,Vaidman2012,Rohrlich2016,Vedral2020}.
In this work we provide an explanation underlying this apparent nonlocality. The role of
confined magnetic field in the Aharonov-Bohm setup in Fig.(\ref{fig:AB}) is to impart
a puncture in the configuration space of the charge, $\mathbb{R}^2$. Therefore, the charged particle responds
only to the modified topology of its configuration space, which is $\mathbb{R}^2-\{0\}$. We find that
quantizing the modified configuration space, $\mathbb{R}^2-\{0\}$, leads to a family of unitarily
inequivalent operator representations, which are parametrized by a real number $\beta$. If $\beta$
is chosen to be $\beta=\alpha \equiv -q\Phi/2\pi$, then the role of charge, $q$, and the magnetic
flux, $\Phi$, is to choose a particular unitary operator representation in the Hilbert space,
${\mathscr H}$, corresponding to the charged particle in $\mathbb{R}^2-\{0\}$.
%%%%%%%%%%%%%%%%%%%%%%%%%%%%%%%%%%%%%%%%%%%%%%%%%%%%%%%%%%%%%%%%%%%%%%%%%%%%%%%%%%%%%%%%%%%%%%%%%%%%%%%
\section{Supplemental Material: Group theoretic quantization of punctured plane}
Let us now quantize the punctured plane using group theoretic quantization procedure.
Let the punctured plane, $X = \R^2-\{0\}$, be the configuration space corresponding
to a classical system. Since the punctured plane is a submanifold of $\R^2$, and is
diffeomorphic to $S^1 \times \R^+$, the punctured plane, $X = \R^2-\{0\}$, becomes
a multiply-connected space with fundamental group $\Z$.
The phase space, \ie, the cotangent bundle, of the punctured plane, $X$, can be straightforwardly
found to be $M = T^*X = X \times \R^2$. Any point in the phase space, $M$, can be written as
$(x,y,p_x, p_y)$, where $(x,y)\in X$, and $(p_x, p_y)\in \R^2$. The symplectic form, $\omega$,
in the phase space, $M$, is induced from $\R^4$, which can be explicitly written as
\begin{equation}
 \omega = \dd x \wedge \dd p_x + \dd y \wedge \dd p_y.
\end{equation}
Quantizing the phase space, $M=X \times \R^2$, corresponding to the punctured plane,
$X = \R^2-\{0\}$, using group theoretic quantization procedure is carried out as shown below:
\vskip 10pt\noindent
$\bullet$ \textbf{Step-I}: Consider the canonical group, $\mathscr{G}$,
as $\mathscr{G} = \R^2 \rtimes (SO(2)\times \R^+)$, with the group product given by
\begin{align}
 (\vb{u},A_\t,\lambda)(\vb{u}', A_{\t'}, \lambda')
 = (\vb{u}+ \lambda^{-1}A_\t\vb{u}', A_{\t+\t'}, \lambda \lambda'),
\end{align}
where $A_{\t}$ is
\begin{equation}
 A_{\t} \equiv
 \begin{pmatrix}
  \cos\t & -\sin\t\\
  \sin\t & \cos\t
 \end{pmatrix}.
\end{equation}
Note that this group product is nothing but the semidirect product induced by the homomorphism,
$\Phi: SO(2) \times \R^+ \to \text{Aut}(\R^2)$, which maps $(A, \lambda) \mapsto \lambda^{-1}A$. Since the semidirect
product of two Lie groups is also a Lie group, one can conclude that the canonical group,
$\mathscr{G}=\R^2 \rtimes (SO(2)\times \R^+)$, is a Lie group as both $\R^2$ and $SO(2)\times \R^+$ are Lie groups.

We consider the action of the canonical group, $\mathscr{G} = \R^2 \rtimes (SO(2)\times \R^+)$,
on the phase space, $M=X \times \R^2$, as
\begin{equation}
 (\vb{u}, A_\t, \lambda)\cdot (\vb{x}, \vb{p}) = (\lambda A_\t \vb{x}, \lambda^{-1} A_\t \vb{p} - \vb{u}),
\end{equation}
and it can be shown straightforwardly that the group action is transitive, and preserves the symplectic form,
$\omega$, corresponding to the phase space, $M=X \times \R^2$. Furthermore, the Lie algebra,
$\mathfrak{g}$, associated with
the canonical group, $\mathscr{G} = \R^2 \rtimes (SO(2)\times \R^+)$,
is a four dimensional algebra, whose elements are represented by
$(b_1,b_2,\t,r) \in \mathfrak{g}$. The basis elements of the Lie algebra, $\mathfrak{g}$, is related to the elements
of the canonical group, $\mathscr{G} = \R^2 \rtimes (SO(2)\times \R^+)$,
through the exponential map, $\exp: \mathfrak{g} \to \mathscr{G}$, as
\begin{align}
 \exp(b_1,0,0,0) &= (b_1,0,I_2,1), \\
 \exp(0,b_2,0,0) &= (0,b_2,I_2,1), \\
 \exp(0,0,\t,0) &= (0,0,A_\t,1), \\
 \exp(0,0,0,r) &= (0,0,I_2,e^r).
\end{align}
Employing the following identity:
\begin{align}\label{bkh}
 (e^{tA}e^{sB})(e^{-tA}e^{-sB}) = e^{ts[A,B] + \text{higher order terms}},
\end{align}
the Lie bracket on the Lie algebra, $\mathfrak{g}$, is determined to be
\begin{widetext}
\begin{align}
 [(b_1, b_2, \t, r),(b'_1,b'_2,\t',r')]
 = \l(\t' b_2 - \t b'_2 + r'b_1 - rb'_1, \t b'_1 - \t' b_1 + r'b_2 - rb'_2, 0, 0\r).
\end{align}
\vskip 10pt\noindent
$\bullet$ \textbf{Step-II}: Since the fundamental vector fields corresponding to the elements
of the Lie algebra, $\mathfrak{g}$, is
\begin{equation}
 \gamma^{(\vb{b},\t,r)}_{(\vb{x},\vb{p})}(f) = \dv{f(\exp(-t(\vb{b},\t,r))(\vb{x},\vb{p}))}{t}
 \bigg|_{t=0},\ \forall ~ f\in C^\infty( M),
\end{equation}
we find the fundamental vector fields in the phase space, $M$, as
\begin{align}
 \gamma^{(\vb{b}, \t, r)} &= (\t y -r x) \pdv{}{x} + (-\t x- r y) \pdv{}{y}
 + (r p_x + \t p_y + b_1) \pdv{}{p_x} + (r p_y - \t p_x + b_2)\pdv{}{p_y}.
\end{align}
\end{widetext}
\vskip 10pt\noindent
$\bullet$ \textbf{Step-III}: In order to show that the fundamental vector fields, $\gamma^{(\vb{b},\t,r)}$, are
Hamiltonian vector fields, one has to construct an observable, say, $f\in C^\infty(M)$, such that
$\dd f(\cdot) = \omega(\gamma^{(\vb{b},\t,r)},\cdot)$. For this purpose, we consider the classical observable
\begin{align}
 f(x,y,p_x,p_y) = -(rxp_x + ryp_y + b_1 x+ b_2 y -\t yp_x + \t xp_y),
\end{align}
which evidently satisfies the required property, and this proves that the vector field, $\gamma^{(\vb{b},\t,r)}$,
is a Hamiltonian vector field. Employing the explicit form of the classical observables, $f$, we define the map,
$P:\mathfrak{g}\to C^\infty(M)$, as
\begin{align}
 P_{(\vb{b}, \t, r)} = r\vb{x}\cdot \vb{p} + \vb{b}\cdot \vb{x} + \t\vb{x}\cdot C \vb{p},
\end{align}
where
\begin{align*}
 C \equiv
 \begin{pmatrix}
  0 & 1\\
  -1 & 0
 \end{pmatrix},
\end{align*} 
so that 
\begin{align}
 \xi_{P_{(\vb{b},\t,r)}} = -\gamma^{(\vb{b},\t,r)}. 
\end{align}
\vskip 10pt\noindent
$\bullet$ \textbf{Step-IV}: The action of the canonical group,
$\mathscr{G} = \R^2 \rtimes (SO(2)\times \R^+)$, on the phase space, $M=X \times \R^2$, is effective,
since $(\lambda A_\t \vb{x}, \lambda^{-1}A_\t\vb{p}-\vb{u}) = (\vb{x}, \vb{p})$, if and only if
$(A_\t - \lambda^{-1}I_2)\vb{x} = 0$ and $(A_\t-\lambda I_2)\vb{p} = \vb{u}$. Since the matrix, $A_\t$, has no
positive real eigenvalues, unless $\t=0$, in which case the eigenvalue is $1$, so that 
$A_\t = I_2$, $\lambda=1$,
and $\vb{u}=0$. Furthermore, the effectiveness of the action
of the canonical group, $\mathscr{G} = \R^2 \rtimes (SO(2)\times \R^+)$, on the phase space,
$M=X \times \R^2$, implies that the map, $\gamma: \mathfrak{g} \to \text{HamVF}(M)$, is one-one.
\vskip 10pt\noindent
$\bullet$ \textbf{Step-V}: The map $P:\mathfrak{g}\to C^\infty(M)$ is a homomorphism, if
$\{P_{(\vb{b}, \t,r)},P_{(\vb{b}', \t',r')} \} = P_{[(\vb{b}, \t, r),(\vb{b}', \t', r)]}$,
which can be verified as below:
\begin{widetext}
\begin{align}
 \{P_{(\vb{b}, \t,r)},P_{(\vb{b}', \t',r')} \}
 &= \pdv{P_{(\vb{b}, \t,r)}}{\vb{x}} \pdv{P_{(\vb{b}', \t',r')}}{\vb{p}}
 - \pdv{P_{(\vb{b}, \t,r)}}{\vb{p}}\pdv{P_{(\vb{b}', \t',r')}}{\vb{x}}, \nonumber\\
 &= \l[(r\vb{p} + \vb{b} + \t C\vb{p})\cdot(r'\vb{x} - \t' C\vb{x})\r]
 - \l[(r'\vb{p} + \vb{b}' + \t' C\vb{p})\cdot(r\vb{x} - \t C\vb{x})\r], \nonumber\\
 &= \vb{x}\cdot \l(r'\vb{b} - r\vb{b} - \t C\vb{b} + \t'C\vb{b}'\r)
 = P_{(\t' b_2 - \t b'_2 + r'b_1 -rb'_1, \t b'_1 - \t b_1 + r'b_2 - rb'_2)}, \nonumber\\
 &= P_{[(\vb{b}, \t, r),(\vb{b}', \t', r)]}.
 \end{align}
\end{widetext}
Therefore, it is evident that the map, $P:\mathfrak{g}\to C^\infty(M)$, is a momentum map.
\vskip 10pt\noindent
$\bullet$ \textbf{Step-VI}: Let $\mathcal{U}:\mathscr{G}\to \text{Unitary}(\mathscr{H})$ be an irreducible
unitary representation of the canonical group, $\mathscr{G} = \R^2 \rtimes (SO(2)\times \R^+)$, which maps
$(\vb{u}, A_\t, \lambda) \mapsto \mathcal{U}(\vb{u}, A_\t, \lambda)$. Defining the operators,
\begin{align}
 U(\t, r) &= \mathcal{U}(\exp(0, \t, r)), \\
 V(\vb{b}) &= \mathcal{U}(\exp(\vb{b}, 0, 0)),
\end{align}
where $r = \log(\lambda)$, we obtain the Weyl-like relations as
\begin{align}\label{eq:weyl-like}
 \begin{split}
  U(\t,r)U(\t',r')&= U(\t+ \t', r+r'),\\
  V(\vb{b})V(\vb{b}') &=V(\vb{b} + \vb{b}'),\\
  U(\t, r)V(\vb{b}) &=V(e^{-r}A_\t\vb{b})U(\t,r).
 \end{split}
\end{align}
Since the operators, $U$ and $V$, can be further written as a product of unitary operators, we essentially
have four one parameter group of unitary operators. Using Stone's theorem \cite{Prugovecki2003} on each of
them, we write
\begin{align}
 U(\t, r) = e^{-i(\t\hat{\pi}_1 + r\hat{\pi}_2)/\hbar} \And V(\vb{b}) = e^{-i(b_1 \hat{c} + b_2 \hat{s})/\hbar}.
\end{align}
Since $SO(2)\times \R^+$ and $\R^2$ are both Abelian groups, it follows that
\begin{align}
 [\hat{\pi}_1, \hat{\pi}_2] = 0, ~~ \text{and} ~~ [\hat{c}, \hat{s}] = 0.
\end{align}
Furthermore, employing the identity in Eq. \eqref{bkh}, we find the commutation relations as
\begin{align}
 [\hat{s}, \hat{\pi}_1] &= i\hat{c}, ~~~~ [\hat{c}, \hat{\pi}_1] = -i\hat{s}\label{com1}, \\
 [\hat{s}, \hat{\pi}_2] &= i\hat{s}, ~~~~ [\hat{c}, \hat{\pi}_2] = i\hat{c}\label{com2}.
\end{align}
Therefore, the quantization map for the phase space, $M=X \times \R^2$,
with the canonical group, $\mathscr{G} = \R^2 \rtimes (SO(2)\times \R^+)$, is given by
\begin{align}
 P^{(\vb{b},0,0)} = b_1 x + b_2 y \mapsto -i(b_1 \hat{c} + b_2\hat{s}),
\end{align}
and
\begin{align}
 P^{(0,\t,r)} &= rxp_x + ryp_y - \t yp_x + \t x p_y \\
 & \mapsto -i(\t\hat{\pi}_1 + r\hat{\pi}_2).
\end{align}
     
Let $\mathscr{H}$ be the space of all square integrable functions on $X=\mathbb{R}^2-\{0\}$
with respect to the measure
$\dd \mu = \dd \phi\dd\rho/(2\pi\rho)$. A weakly continuous, irreducible unitary representation of the
canonical group, $\mathscr{G}$, in $\text{Unitary}(L^2(X))$ can be written as
\begin{subequations}\label{eq:reps}
\begin{align}
 U(\t, \lambda) \psi(\p, \rho) &= \psi((\p-\t)\text{mod}(2\pi), \lambda^{-1}\rho), \label{eq:repa}\\
 V(\vb{b}) \psi(\p, \rho) &= e^{-i(b_1\cos\p + b_2\sin\p)\rho/\hbar} \psi(\p, \rho),\label{eq:repb}
\end{align}
\end{subequations}
where $\psi\in L^2(X)$. Therefore, the representation of the self-adjoint operators can be deduced by
expanding both sides of the Eqs. \eqref{eq:repa} and \eqref{eq:repb} as a power series as
\begin{align}
 \hat{c} \psi(\p, \rho) &= (\rho \cos\p) \psi(\p, \rho), \\
 \hat{s} \psi(\p, \rho) &= (\rho \sin\p) \psi(\p, \rho),
\end{align}
and
\begin{align}
 \hat{\pi}_1 \psi(\p, \rho) &= -i\hbar\pdv{\psi}{\phi}, \\
 \hat{\pi}_2 \psi(\p, \rho) &= -i \hbar\rho \pdv{\psi}{\rho}.
\end{align}
    
If the canonical group,
$\mathscr{G} = \R^2 \rtimes (SO(2)\times \R^+)$, admits universal
covering groups, then the concerned covering group could also be considered as a new canonical group associated
with the phase space, $M$ \cite{isham}. Consider the universal covering group,
$\tilde{\mathscr{G}} = \R^2 \rtimes (\R\times \R^+)$, with a universal covering map,
$\pi:\tilde{\mathscr{G}} \to \mathscr{G}$, that has the group product
\begin{align}
 (\vb{u}, \t, \lambda)\cdot (\vb{u}', \t', \lambda')
 = (\vb{u} + \lambda^{-1} A_{\t}\vb{u}', \t+\t', \lambda\lambda'),
\end{align}
and $\pi(\vb{u},\t,\lambda)= (\vb{u}, A_\t, \lambda)$. The action of the universal covering group,
$\tilde{\mathscr{G}}$, on the phase space, $M$, is induced from the action of the base group,
$\mathscr{G} = \R^2 \rtimes (SO(2)\times \R^+)$, on
the phase space, $M$, \ie $\ \tilde{g}\cdot (\vb{x},\vb{p}) = \pi(g)\cdot x$, where $\tilde{g}\in \tilde{\mathscr{G}}$.
The covering group, $\tilde{\mathscr{G}}$, acts effectively everywhere, except for the discrete subgroup $2\pi \Z$ of
$\tilde{\mathscr{G}}$. Employing the twisted representations \cite{isham}, we find a family of inequivalent
unitary representations, parametrized by $\alpha \in \R$, of the new
canonical group, $\tilde{\mathscr{G}}$, which are given by
\begin{subequations}\label{neq_rep}
\begin{align}
 U(\t, \lambda) \psi(\p, \rho) = e^{-i\a \t} \psi((\p -\t)\text{mod}(2\pi), \lambda^{-1}\rho), \\
 V(\vb{b}) \psi(\p, \rho) = e^{-i(b_1\cos(\p) + b_2\sin(\p))\rho/\hbar} \psi(\p, \rho),
\end{align}
\end{subequations}
and which results in an operator representation as
\begin{align}\label{eq:operators}
 \hat{\pi}_1 \psi(\p, \rho) &= -i\hbar\pdv{\psi}{\phi} + \hbar\a \psi, \\
 \hat{\pi}_2 \psi(\p, \rho) &= -i \hbar\rho \pdv{\psi}{\rho}, \\
 \hat{c}\psi(\phi,\rho) &= \rho \cos(\phi)\psi(\phi,\rho), \\
 \hat{s}\psi(\phi,\rho) &= \rho \sin(\phi)\psi(\phi,\rho).
\end{align}
\vskip 5pt
%\begin{acknowledgments}
%Acknowledgments: 
%\end{acknowledgments}
%%%%%%%%%%%%%%%%%%%%%%%%%%%%%%%%%%%%%%%%%%%%%%%%%%%%%%%%%%%%%%%%%%%%%%%%%%%%%%%%%%%%%%%%%%%%%%%%%%%%%%%

%%%%%%%%%%%%%%%%%%%%%%%%%%%%%%%%%%%%%%%%%%%%%%%%%%%%%%%%%%%%%%%%%%%%%%%%%%%%%%%
\end{document}